\documentclass[10pt,conference]{IEEEtran}
\usepackage{cite}
\usepackage{amsmath,amssymb,amsfonts}
\usepackage{graphicx}
\usepackage{textcomp}
\usepackage[usenames,dvipsnames]{xcolor}
\usepackage{algorithm}
\usepackage[noend]{algpseudocode}
\usepackage[framemethod=default]{mdframed}
\usepackage{comment}
\usepackage{fancyhdr}
\usepackage{wrapfig}
\usepackage{listings}
\usepackage{subcaption}
\usepackage[hyphens]{url}
\usepackage{array,tabularx}
\usepackage[most]{tcolorbox}
\usepackage[ddmmyyyy]{datetime}
\usepackage[hyperfootnotes=false]{hyperref}
\usepackage{cleveref}
\usepackage{soul}
\usepackage{enumitem}
\usepackage{balance}
\usepackage{setspace}
\usepackage{textcomp}

\def\BibTeX{{\rm B\kern-.05em{\sc i\kern-.025em b}\kern-.08em
    T\kern-.1667em\lower.7ex\hbox{E}\kern-.125emX}}

\crefformat{section}{\S#2#1#3} 
\crefformat{subsection}{\S#2#1#3}
\crefformat{subsubsection}{\S#2#1#3}

\newmdenv[linecolor=blue,backgroundcolor=cyan,linewidth=2pt,  innertopmargin=0pt,
    innerbottommargin=0pt,
    innerleftmargin=0cm,
    innerrightmargin=0cm]{myframe}

\tcbset{
    frame code={}
    center title,
    left=0pt,
    right=0pt,
    top=0pt,
    bottom=0pt,
    colback=gray!35,
    colframe=white,
    width=\dimexpr\linewidth\relax,
    enlarge left by=-1mm,
    enlarge right by=-1mm,
    boxsep=3pt,
    arc=0pt,outer arc=0pt,
    }

\definecolor{mGreen}{rgb}{0,0.6,0}
\definecolor{mGray}{rgb}{0.5,0.5,0.5}
\definecolor{mPurple}{rgb}{0.58,0,0.82}
\definecolor{backgroundColour}{rgb}{0.85,0.85,0.82}
\definecolor{mygray}{gray}{0.85}
\definecolor{grey}{cmyk}{0,0,0,.1}

\lstdefinestyle{CStyle}{
    backgroundcolor=\color{backgroundColour},
    commentstyle=\color{mGreen},
    keywordstyle=\color{magenta},
    numberstyle=\tiny\color{mGray},
    stringstyle=\color{mPurple},
    basicstyle=\footnotesize,
    breakatwhitespace=false,
    breaklines=true,
    captionpos=b,
    keepspaces=true,
    numbers=left,
    numbersep=5pt,
    showspaces=false,
    showstringspaces=false,
    showtabs=false,
    tabsize=2,
    escapechar=\%,
    language=C
}

\newcommand\blfootnote[1]{%
  \begingroup
  \renewcommand\thefootnote{}\footnote{#1}%
  \addtocounter{footnote}{-1}%
  \endgroup
}

\newcommand{\squishenumstart}[1][1)]{
 \begin{enumerate}[#1]
  { \setlength{\itemsep}{0pt}
     \setlength{\parsep}{0pt}
     \setlength{\topsep}{3pt}
     \setlength{\partopsep}{0pt}
     \setlength{\leftmargin}{1em}
     \setlength{\labelwidth}{1em}
     \setlength{\labelsep}{0.5em} } }

\newcommand{\squishenumend}{
  \end{enumerate}  }

\newcommand{\squishlist}{
 \begin{list}{$\circ$}
  { \setlength{\itemsep}{0pt}
     \setlength{\parsep}{0pt}
     \setlength{\topsep}{3pt}
     \setlength{\partopsep}{0pt}
     \setlength{\leftmargin}{1em}
     \setlength{\labelwidth}{1em}
     \setlength{\labelsep}{0.5em} } }

\newcommand{\squishend}{
  \end{list}  }

\makeatletter
\g@addto@macro{\normalsize}{%
  \setlength{\abovedisplayskip}{3pt plus 0.5pt minus 1pt}
  \setlength{\belowdisplayskip}{3pt plus 0.5pt minus 1pt}
  \setlength{\abovedisplayshortskip}{0pt}
  \setlength{\belowdisplayshortskip}{0pt}
  \setlength{\intextsep}{4pt plus 1pt minus 1pt}
  \setlength{\textfloatsep}{4pt plus 1pt minus 1pt}
  \setlength{\skip\footins}{5pt plus 1pt minus 1pt}}
\makeatother

\makeatletter
\algrenewcommand\ALG@beginalgorithmic{\footnotesize}
\makeatother

\algrenewcommand{\algorithmiccomment}[1]{\hfill$\triangleright$ \textit{#1}}
\algrenewcommand\algorithmicindent{1.0em}

\algnewcommand{\IIf}[1]{\State\algorithmicif\ #1\ \algorithmicthen}
\algnewcommand{\EndIIf}{\unskip\ \ }

\newcommand{\myName}{MATSA}

\title{\vspace{-16pt}Accelerating Time Series Analysis via \\
Processing using Non-Volatile Memories\vspace{-16pt}}
\newcommand{\affilUMA}[0]{\textsuperscript{\S}}
\newcommand{\affilBSC}[0]{\textsuperscript{\P}}
\newcommand{\affilETH}[0]{\textsuperscript{$\dagger$}}
\newcommand{\affilNTUA}[0]{\textsuperscript{$\ddagger$}}

\author{
\hspace{-12pt}
\fontsize{11.4}{8}\selectfont
\parbox[t]{1.02\textwidth}{
{Ivan Fernandez\affilUMA\affilETH\affilBSC}\hspace{6pt}
{*Christina Giannoula\affilETH\affilNTUA}\hspace{6pt}
{*Aditya Manglik\affilETH}\hspace{6pt}
{Ricardo Quislant\affilUMA}\hspace{6pt}
{Nika Mansouri Ghiasi\affilETH}
{\vspace{-1pt}\textcolor{white}{}}
\\
{\vspace{3pt}\textcolor{white}{s}}\hspace{38pt}
{Juan Gómez-Luna\affilETH}\hspace{10pt}
{Eladio Gutierrez\affilUMA}\hspace{10pt} 
{Oscar Plata\affilUMA}\hspace{10pt}%
{Onur Mutlu\affilETH} 
\\
\vspace{-7pt}
\centering\small{\emph{{\affilUMA University of Malaga  \hspace{10pt}  \affilETH ETH Z{\"u}rich \hspace{10pt} \affilBSC Barcelona Supercomputing Center \hspace{10pt} \affilNTUA  National Technical University of Athens
}}}%
}
\vspace{-4pt}
}

\begin{document}
\sloppy
\maketitle
\thispagestyle{plain}
\pagestyle{plain}

\begin{abstract}
\emph{Time Series Analysis} (\emph{TSA}) is a critical workload to extract valuable information from collections of sequential data, e.g., detecting anomalies in electrocardiograms. Subsequence Dynamic Time Warping (sDTW) is the state-of-the-art algorithm for high-accuracy TSA. We find that the performance and energy efficiency of sDTW on conventional CPU and GPU platforms are heavily burdened by the latency and energy overheads of data movement between the compute and the memory units. sDTW exhibits low arithmetic intensity and low data reuse on conventional platforms, stemming from poor amortization of the data movement overheads. To improve the performance and energy efficiency of the sDTW algorithm, we propose MATSA, the first \underline{M}agnetoresistive RAM (MRAM)-based \underline{A}ccelerator for \underline{TSA}. MATSA leverages Processing-Using-Memory (PUM) based on MRAM crossbars to minimize data movement overheads and exploit parallelism in sDTW. MATSA improves performance by 7.35$\times$/6.15$\times$/6.31$\times$ and energy efficiency by 11.29$\times$/4.21$\times$/2.65$\times$ over server-class CPU, GPU, and Processing-Near-Memory platforms, respectively.
\blfootnote{\textbf{* Christina Giannoula and Aditya Manglik have equal contribution.}}
\end{abstract}
\section{Introduction}
In the era of Internet-Of-Things and Big Data, emerging applications operate on petabyte-scale datasets that are increasingly difficult to store and analyze. Small sensors and edge devices continuously generate data sampled over time, resulting in time-ordered observations (e.g., temperature or voltage). Such a collection of data values is referred to as a \emph{time series} (TS)~\cite{esling2012time}. TS is a common data representation in many real-world scientific applications, including sensing, genomics, neuroscience, financial markets, epidemiology, and environmental sciences~\cite{mueen2016extracting}. 

Time series analysis (TSA) splits the time series into \textit{subsequences} of consecutive data points to extract valuable information from large datasets. This information can help filter relevant subsequences to minimize the cost of applying complex and expensive domain-specific analysis algorithms. A real-life example is the detection of anomalies in an electrocardiogram and the elimination of subsequences that indicate normal behavior~\cite{yao2022modified}. TSA determines subsequences of interest using different similarity approaches, such as the Euclidean Distance (ED) or the subsequence Dynamic Time Warping (sDTW). Prior work demonstrates that sDTW provides a higher precision than ED in most scenarios~\cite{alaee2021time}; as such, we focus on optimizing sDTW algorithm for TSA analysis. 

sDTW is an embarrassingly parallel workload, because each query can be executed without data dependencies from other queries by multiple concurrent processing units. However, sDTW builds a 2D dynamic programming matrix that incurs quadratic runtime and memory complexity. To understand the bottlenecks of sDTW in state-of-the-art conventional CPU and GPU architectures, we comprehensively characterize the kernel's performance on these platforms (\cref{sec:bottlenecks}). We observe significant performance and energy efficiency overheads in sDTW due to: 1) underutilization of the execution units, and 2) a large number of expensive main memory accesses. The first problem stems from the low number of operations that the sDTW kernel executes per byte brought from memory, which keeps the arithmetic units idle for the largest part of the execution time. The second problem stems from the large memory footprint of the dynamic programming matrix, causing poor spatial and temporal locality. Consequently, sDTW exhibits poor performance on CPU and GPU platforms.

To overcome the memory access challenge, prior works~\cite{gomez2021benchmarking, giannoula2022sparsep,mutlu2022modern} have considered \textit{memory-centric} platforms that integrate processing and storage elements on the same chip to reduce data movement across the constrained data bus that connects a CPU to main memory~\cite{mutlu2019processing, boroumand2018google}. Based on that, we implement and characterize sDTW in a real Processing-Near-Memory (PNM) platform, \textit{UPMEM}~\cite{devaux2019true}, and observe that this new platform does \emph{not} provide performance benefits compared to CPU and GPU executions, due to the large latency of simple operations such as addition and comparison operators. Overall, we conclude that the sDTW kernel exhibits memory-bound behavior on CPU and GPU platforms and compute-bound behavior on the PNM platform (\cref{sec:bottlenecks}). 

In contrast to PNM, \textit{Processing-Using-Memory} (PUM)~\cite{angizi2019aligns,li2017drisa,angizi2019mrima,angizi2018cmp, seshadri2017ambit,mutlu2022modern} executes operations using the memory cells and sense amplifiers, completely eliminating the memory and compute dichotomy. PUM enables 1) performing computation \emph{in the memory array}, since the memory units that store the data also execute the computation, and 2) exploiting a much larger amount of parallelism available in the memory microarchitectures (
as high as the number of crossbar columns available~\cite{hajinazar2021simdram}, i.e., thousands) compared to conventional CPU and GPU systems. From the technology perspective, non-volatile memories (NVM) offer a promising substrate to implement PUM~\cite{roy2020memory}. However, different NVM substrates exhibit varying latency, energy, and endurance characteristics, a key design constraint for different accelerators~\cite{perez2019toward}. Magnetoresistive RAM (MRAM)-based PUM substrates offer low read/write latencies, low energy per operation, and high endurance~\cite{daulby2020comparing, lin2022all}. Considering these characteristics, we explore MRAM as a potential NVM substrate to accelerate the sDTW kernel.

To this end, our goal in this work is to leverage MRAM-based PUM to \emph{enable high-performance and energy-efficient sDTW execution for a wide range of applications}. We propose \emph{\myName{}}, the first \underline{M}RAM-based \underline{A}ccelerator for \underline{TSA}. \myName{} derives its performance benefits from three key mechanisms. First, \myName{} decomposes sDTW's computational kernel into simple bitwise boolean computations and executes them in the MRAM crossbar. This key idea significantly minimizes data movement overheads as it is performed where data resides. Second, we implement a novel data mapping that reduces the runtime memory footprint of sDTW from quadratic to linear based on four vectors. 
This key idea enables computing the complete 2D dynamic programming matrix on-the-fly without storing it. Third, \myName{} integrates an effective computation scheme 
that overcomes the inter-cell computation dependencies of the matrix by 1) following an anti-diagonal approach and 2) exploiting pipelining to increase parallelism.

We evaluate \myName{}'s performance based on state-of-the-art latency and energy characteristics of MRAM devices~\cite{yu2016emerging, gallagher2019recent}. To do so, we  implement an in-house simulator for \myName{} and select 64 synthetic datasets to understand its design tradeoffs. Then, we use six real-world datasets (\textit{Human}, \textit{Song}, \textit{Penguin}, \textit{Seismology}, \textit{Power} and \textit{ECG}) to compare three different versions of \myName{} against other state-of-the-art platforms, showcasing its applicability to a wide range of real case scenarios. Our evaluation shows that \myName{} improves performance by 7.35$\times$/6.15$\times$/6.31$\times$ and energy efficiency by 11.29$\times$/4.21$\times$/2.65$\times$ over server-class CPU, GPU, and PNM platforms, respectively.


In summary, we make the following novel contributions:

\begin{itemize}
    \item We thoroughly characterize the state-of-the-art sDTW time series analysis (TSA) algorithm's  performance and energy efficiency on conventional CPU, GPU, and PNM (UPMEM) platforms. Our characterization leads to new observations about the characteristics of sDTW that limit its acceleration in current conventional hardware.

    \item We propose \emph{\myName{}}, the first \underline{M}RAM-based \underline{A}ccelerator for \underline{TSA}. \myName{} 1) exploits  a novel data mapping tailored for MRAM substrates that reduce memory footprint in sDTW, 2) efficiently performs computation in-memory to avoid off-chip data movement, and 3) provides an effective computation scheme to increase parallelism.
    
    \item We conduct a comprehensive evaluation of \myName{} across a diverse set of synthetic and real-world datasets. Our results showcase 6.60$\times$ average improvement in overall performance and a average 6.05$\times$ boost in energy efficiency over state-of-the-art compute-centric and memory-centric platforms.
\end{itemize}
\section{Background \& Motivation}

\subsection{Time Series Analysis}
A {\em time series} $T$ is a sequence of $n$ data points $t_{i}$, where $1\leq i\leq n$, collected over time. A subsequence of $T$, also known as a
\textit{window}, is denoted by $T_{i,m}$, where $i$ is the index of the first data point, and $m$ is the number of samples in the subsequence, with $1\leq i$, and $m\leq n - i$.

There are two main approaches to perform time series analysis: 1) the self-join, and 2) the query-filtering. In self-join, all sequences of a given time series are compared against the remaining subsequences of the same time series. In contrast, query filtering compares a set of queries against a reference.

Time series analysis algorithms usually define a distance metric to measure the similarity between two subsequences. Based on such distance metric, the literature classifies the subsequences with low distance as \textit{motifs}~\cite{patel2002mining} (similarities) and high distance as \textit{discords}~\cite{keogh2007finding} (anomalies). The state-of-the-art set of tools to perform time series analysis is Matrix Profile~\cite{MPROFILEI} (MP). Due to lower computation requirements, prior MP algorithms utilize one-to-one Euclidean Distance as the similarity metric. Recent proposals~\cite{alaee2021time} have started to utilize Dynamic Time Warping (DTW)-based solutions because of higher precision~\cite{ratanamahatana2004making}. DTW enables the detection of events of interest in out-of-sync subsequences, e.g., in subsequences that have different sampling rates. 

\begin{figure}[h!]
    \centering
    \includegraphics[width=\linewidth]{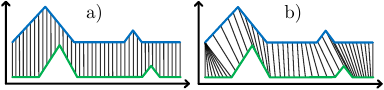}
    \caption{Example of similarity calculation between two subsequences (blue and green). The one-to-one approach in a) provides a low similarity as it only compares each $i^{th}$ point of blue with each $i^{th}$ point of green. In contrast, DTW in b) successfully matches the points of the subsequences.}
    \label{fig:euclid_vs_dtw}
\end{figure}

Figure~\ref{fig:euclid_vs_dtw} shows the key difference between the one-to-one and the DTW approaches, in which we compare two similar-shape subsequences that differ in their offset and scale. 

We observe that the DTW algorithm offers better results as it compares a given point with respect to several potential candidates (i.e., determines the best alignment). In contrast, one-to-one executes point-to-point alignment that cannot determine the best alignment in the presence of an offset. One-to-one can be considered as a special case of DTW where the \textit{warping window} is set to '1'. Therefore, we aim to optimize DTW, a more generic and high-precision algorithm, to provide a TSA accelerator for a wide range of applications.

\subsection{Time Series Analysis Applications}
Time series analysis constitutes one of the most important and general data mining primitives for a wide range of real-world applications~\cite{SS17}: epidemiology, genomics, neuroscience, medicine, environmental sciences, economics, and many more. Table~\ref{tab:tsa_applications} presents a few examples for applications of TSA.

\begin{table}[h!]
\centering
\resizebox{\linewidth}{!}{\begin{tabular}{|l|c||l|c|}
\hline
    \textbf{Field} & \textbf{References} & \textbf{Field} & \textbf{References}\\
\hline\hline
    Bioinformatics     &  \cite{brown2012adict, straume2004dna, bar2004analyzing} & Speech Recognition  &  \cite{balasubramanian2016discovering}\\
\hline
     Robotics          &  \cite{TIU05,Vahdatpour2009toward} & Weather Prediction &  \cite{McGovern2011identifying}\\
\hline
     Neuroscience & \cite{brown2012adict,Kolb1821evidence} & Entomology & \cite{Szigeti2015searching}\\
\hline
     Machine Learning & \cite{WarrenLiao2005clustering,Aghabozorgi2015time,KK03} & Geophysics & \cite{trugman2017growclust,Waldhauser2000adouble,Christophersen2018bayesian,McKee2018instrumental}\\
\hline
    Econometrics & \cite{Howrey1980therole} & Statistics & \cite{Shymway1988applied}\\
\hline
    Finance & \cite{Tsay2005analysis,nakagawa2019stock} & Control Engineering & \cite{George2009time, nelson1998two, barber2011bayesian}\\
\hline
    Signal Processing & \cite{Kumar2012time} & Pattern Recognition & \cite{berlin1994pattern}\\
\hline
    Communication & \cite{lakhina2004characterization, jain1986packet, barford2001characteristics} & Medicine & \cite{Hussain2017symbolic,Balli2008eeg, dunn2021squigglefilter,chen2020anomaly}\\
\hline
    Astronomy & \cite{Vio2004timeseries,Scargle1998studies} & Social Networks & \cite{nusratullah2015detecting, asur2010predicting}\\
\hline
    Clustering & \cite{WarrenLiao2005clustering,Aghabozorgi2015time} & Classification & \cite{KK03}\\
\hline
    Earthquakes & \cite{trugman2017growclust,Waldhauser2000adouble,Christophersen2018bayesian,McKee2018instrumental} & GPS Tracking & \cite{klos2018detecting}\\
\hline
    Virtual Reality & \cite{stoermer2000monitoring} & Gesture Recognition & \cite{hussain2012user,calin2016gesture}\\
\hline
    Trajectories & \cite{ayhan2016time} & Traffic Monitoring & \cite{li2015trend}\\
\hline
\end{tabular}}
\caption{Time Series Analysis main applications}
\label{tab:tsa_applications}
\end{table}

In statistics, econometrics, meteorology, and geophysics, the primary goal of time series analysis is prediction and forecasting. At the same time, in 
signal processing, control engineering, and communication engineering, it is used for signal detection and estimation. In data mining, pattern recognition, and machine learning, time series motif and discord discovery are used for clustering, classification, anomaly detection, and forecasting. Finally, the most important application of time series motif and discord discovery is clustering seismic data and discovering earthquake pattern clusters from the continuous seismic recording. Consequently, seismic clustering can be applied to earthquake relocation and volcano monitoring to help improve earthquake and volcanic hazard assessments.

\begin{figure}[h!]
    \centering
    \includegraphics[width=\linewidth]{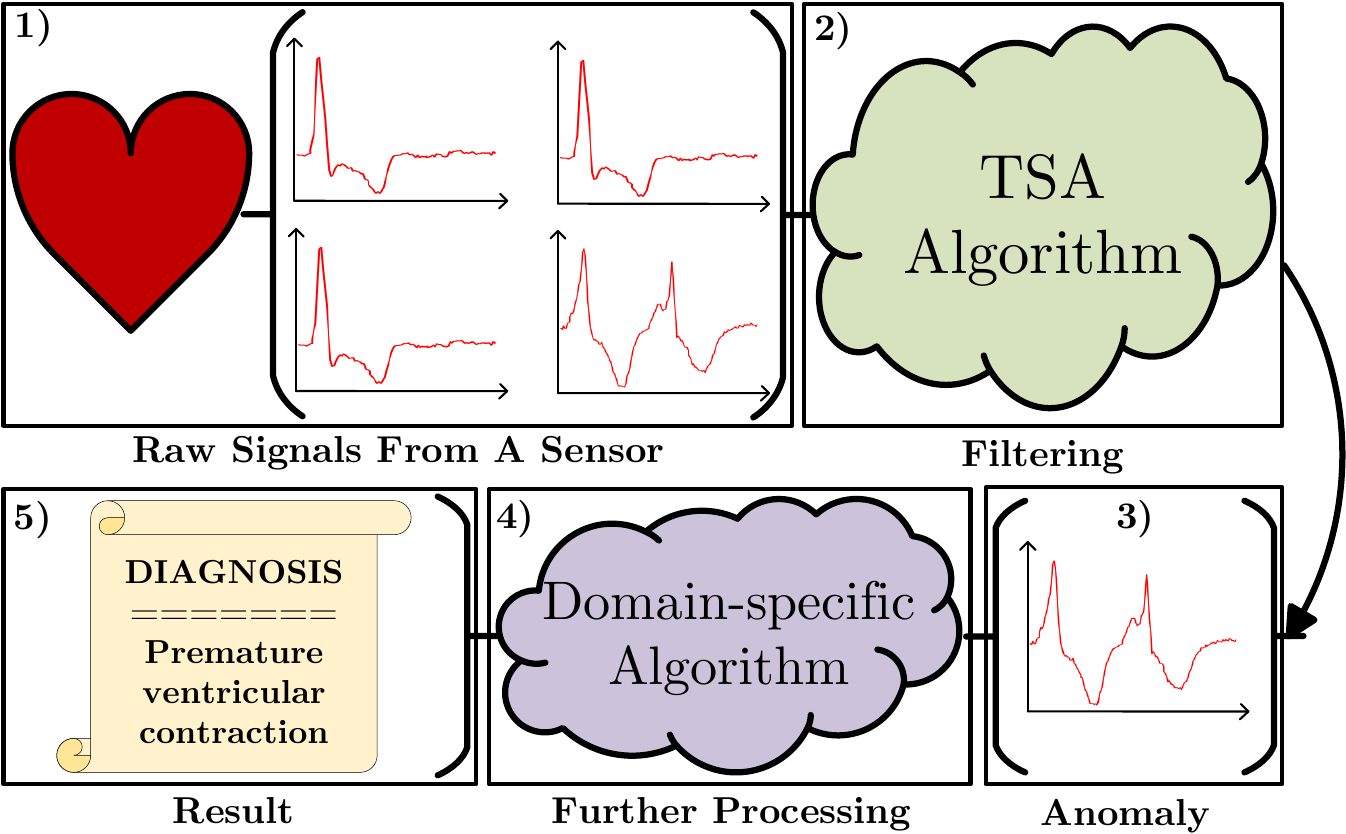}
    \caption{Example TSA application, where TSA acts as a filter to avoid most of the computation. TSA selects the relevant queries (anomalies) and discards the irrelevant ones.}
    \label{fig:tsa_example}
\end{figure}

Within this field, the subsequence Dynamic Time Warping (sDTW) algorithm is a fundamental kernel due to its superior accuracy and generality when compared to other TSA methods~\cite{alaee2021time}. Examples of real-life use cases that can benefit from high-performance and energy-efficient sDTW are:

\begin{itemize}
    \item \textbf{Circulatory Failure Detection in Intensive Care Units}. TSA consumes 90\% of the end-to-end execution time~\cite{hyland2020early}. Figure~\ref{fig:tsa_example} describes the aforementioned process based on an example processing flow. 
           
    \item \textbf{Electroencephalography (ECG)}. TSA is deployed to monitor and filter ECG readings when monitoring patients~\cite{chen2020anomaly}.
    
    \item \textbf{Earthquake Detection}. TSA is critical to process seismograph data and detect anomalies for further analysis~\cite{Christophersen2018bayesian}.
\end{itemize}

\subsection{Dynamic Time Warping (DTW)}
DTW algorithm was first introduced by~\cite{berndt1994using}. The first step of DTW is to compute the distance between a particular point from a subsequence and a set of points from another subsequence, only keeping the minimum of them. This process is repeated for all the points of the first subsequence. Then, DTW computes the addition of all distances, providing a similarity measure between the subsequences (the lower the distance, the higher the similarity).

Assuming that we have two  subsequences, \textit{Q} (query) and \textit{R} (reference), of length n and m, respectively, where:
\begin{equation}
  \begin{array}{l}
    Q = q_{1},q_{2},...,q_{i},...,q_{n} 
    \hspace{5mm}
    R = r_{1},r_{2},...,r_{j},...,r_{m}
  \end{array}   
\end{equation}

DTW constructs an n-by-m scoring matrix (\textit{S}) to determine the similarity between the two subsequences. Each $(i^{th}, j^{th})$ cell of the matrix ($s_{i,j}$) is filled in two steps. First, the algorithm calculates the distance $d(q_{i},r_{j})$ between the two corresponding points of the subsequences. There are several approaches to calculate such distance, while $d(q_{i},c_{j}) = abs(q_{i} - c_{j})$ and $d(q_{i},c_{j}) = (q_{i} - c_{j})^2$ are the most common ones~\cite{berndt1994using}. Second, the distance value is added to the minimum of the three neighboring cells as follows:
\begin{equation}
    s_{i,j} = d(q_{i}, c_{j}) + min(s_{i-1,j-1}, s_{i-1,j}, s_{i,j-1})
\end{equation}
The algorithm fills the entire matrix using dynamic programming. Then, the goal is to find the best alignment (i.e., minimum accumulated cost), known as the \textit{warping path} ($W$). $W$ is a contiguous set of matrix cells that defines the best mapping between \textit{Q} and \textit{R}.

\textbf{Subsequence Dynamic Time Warping (sDTW).} sDTW is a more general DTW algorithm that allows the query to be aligned with part of the reference. Algorithm~\ref{alg:sdtw} presents the pseudocode of sDTW. 

\begin{algorithm}[h!]
    \caption{Subsequence DTW (sDTW)} \label{alg:sdtw}
    \begin{algorithmic}[1]
        \Procedure{sDTW}{Q,R}
        \State $S \leftarrow zeros(N, M);$
        \State $S[0, 0] = dist(Q[0], R[0]);$
        \For {$i \leftarrow 1$ to $N$} 
         \State $S[i,0] \leftarrow S[i-1,0] + dist(Q[i], R[0]);$
        \EndFor
        \For { $i \leftarrow 1$ to $N$} 
            \For {$j \leftarrow 1$ to $M$}
            \State $S[i,j] \leftarrow dist(Q[i], R[j])$ +
            \State $min(S[i-1,j-1], S[i,j-1], S[i-1,j]);$
            \EndFor
        \EndFor
        \Return{$min(S[N,:])$}
        \EndProcedure
    \end{algorithmic}
\end{algorithm}

First, sDTW initializes the matrix \textit{S} with zeros. Second, it calculates the distance value of the top-left corner and then the remaining elements of the first row, taking into account the previous values. Third, it fills the remaining elements of the matrix using dynamic programming row by row. Finally, it returns the minimum element of the last row of the \textit{S} matrix, which indicates the similarity between the query and the best alignment with (part of) the reference. The nested \texttt{for} loops (lines 6 and 7 in Algorithm~\ref{alg:sdtw}) are responsible for the quadratic runtime and memory complexities.

\subsection{Bottlenecks of sDTW in Conventional and PNM Platforms}
\label{sec:bottlenecks}
sDTW's quadratic computational complexity is challenging to overcome, especially when accurate results are required and algorithmic optimizations are insufficient~\cite{wu2020fastdtw}. To determine the bottlenecks in conventional platforms, we perform a detailed characterization of parallelized and optimized sDTW kernels on CPU, GPU, FPGA, and PNM platforms.

\begin{figure}[h!]
    \centering
    \includegraphics[width=\linewidth]{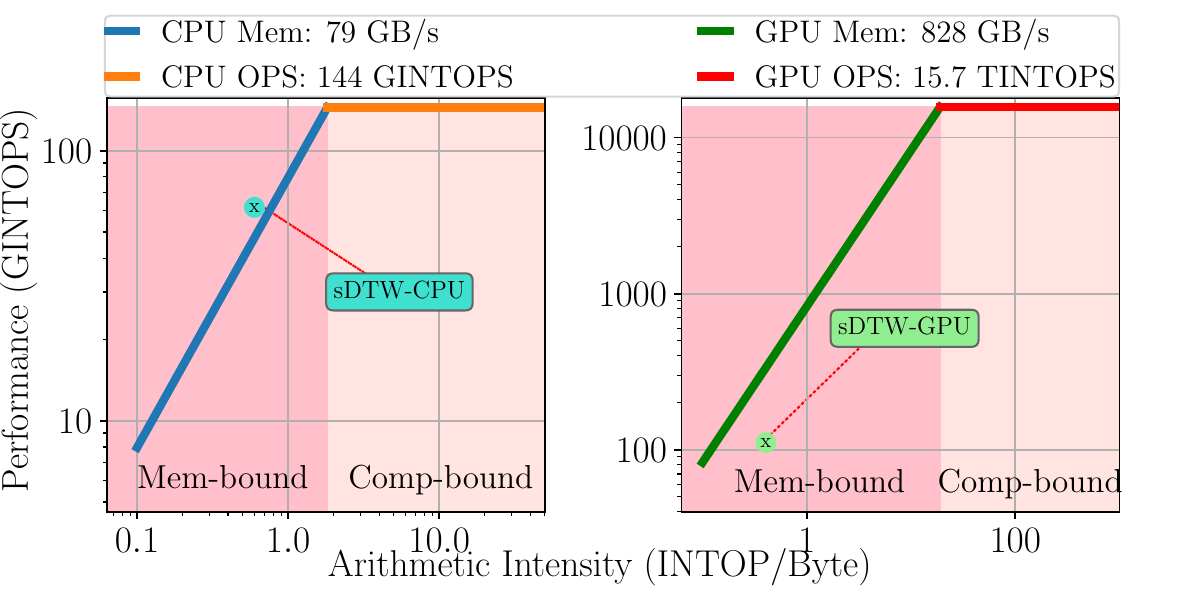}
    \caption{Roofline plots for sDTW on a many-core CPU platform (left) and a server-class GPU (right).}
    \label{fig:sdtw_roofline}
\end{figure}

\textbf{CPU}. We profile the performance of sDTW on an Intel Xeon Phi 7210 CPU using the Intel Advisor tool~\cite{koskela2017roofline}. We build the roofline plot and present the result in Figure~\ref{fig:sdtw_roofline}-left. First, we observe that sDTW-CPU can utilize only 41\% of the system's integer peak performance, i.e., 59 GINTOPS out of 145 GINTOPS, and exhibits low arithmetic intensity (0.55 INTOP/Byte). Second, the total memory traffic generated during runtime is 267 GB. In contrast, the memory footprint of the sDTW kernel is only 570 MB. This demonstrates that sDTW is a memory-bound kernel for CPU targets.

\textbf{GPU}. Several prior works propose accelerating sDTW using GPUs (e.g., \cite{SH20}). However, these implementations are tailored and optimized for specific workload sizes. They rely on high-latency global memory when working with arbitrary-sized datasets, which results in large performance penalties compared to the optimal input size. To quantify the bottlenecks, we develop an optimized CUDA-based implementation that supports arbitrary subsequence sizes and characterize it on the NVIDIA Tesla V100 GPU. We analyze the sDTW kernel using NVIDIA Visual Profiler~\cite{NVVP} and generate the roofline plot in Figure~\ref{fig:sdtw_roofline}-right. We observe that sDTW-GPU's performance improves with respect to sDTW-CPU but utilizes merely 1\% of the GPU's available peak performance. We explain this observation by 1) the low arithmetic intensity of sDTW and 2) the limited per-thread available local memory. Even increasing the available local memory does not improve performance and the algorithm hits the memory roof due to 1), thus greatly underutilizing the platform. Based on this analysis, we conclude that GPU is not a good target for sDTW kernels executing on arbitrary subsequence sizes, which is the common case in many applications.    

\begin{figure}[h!]
    \centering
    \includegraphics[width=0.99\linewidth]{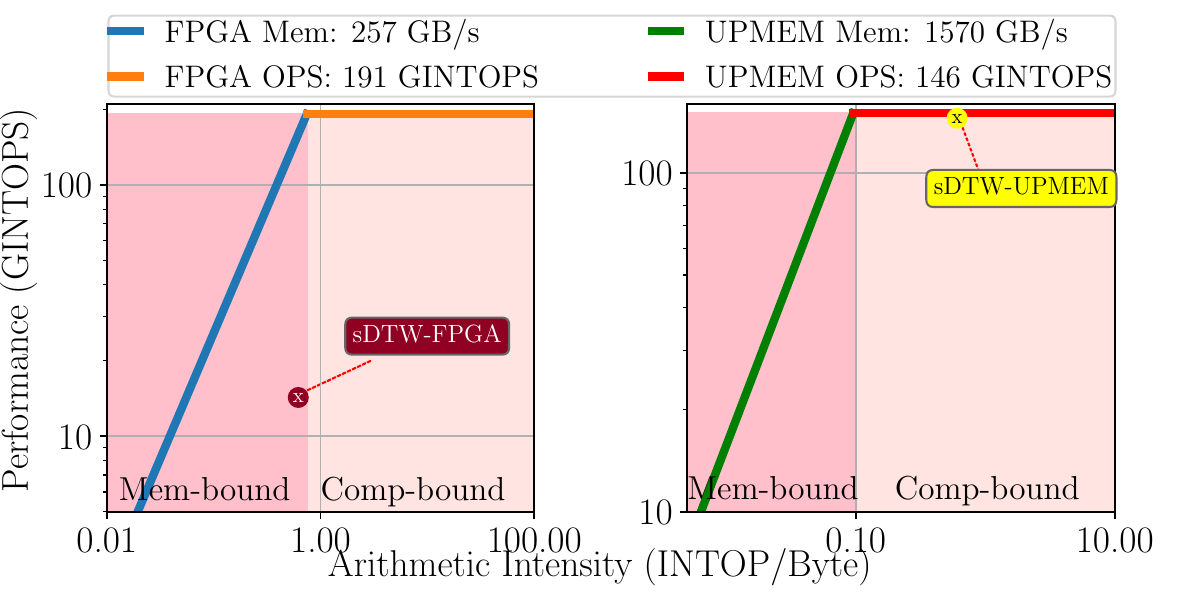}
    \caption{Roofline plots for sDTW on FPGA (left) and UPMEM (right) platforms.}
    \label{fig:sdtw_roofline_upmem}
\end{figure}

\textbf{FPGA}. sDTW acceleration using FPGAs requires large onboard memory to achieve high performance. As most of the prior work based on FPGAs does not provide high on-chip memory capacity, data is distributed over the chip. We develop an optimized FPGA implementation targeting a Xilinx Alveo U50 and build the roofline model in Figure~\ref{fig:sdtw_roofline_upmem}-left. We observe that the eight compute units that fit in the FPGA achieve less than 7\% of the available peak throughput and are insufficient to exploit the inherent parallelism in the sDTW kernel.

\vspace{1mm}
\noindent
\setlength{\fboxsep}{7pt}
\setlength{\fboxrule}{1pt}
\fcolorbox{black}{grey}{%
    \parbox{0.93\linewidth}{%
    \noindent\textbf{Key Observation 1:} Conventional architectures fail to provide a high performance and energy efficient acceleration solution because execution time and energy are wasted on the data movement between memory and processing units.
    }       
}
\vspace{1mm}

\textbf{Processing-Near-Memory (PNM)}. PNM platforms place processing units in the same die as memory units. The idea behind this paradigm is to exploit the lower latency and higher bandwidth available in memory and mitigate the data movement overheads between the processing units and memory. To evaluate the performance and energy efficiency of sDTW on PNM, we implement an optimized version of the algorithm on UPMEM~\cite{upmem2018}, the first commercially available server-class PNM platform. We build the roofline model in Figure~\ref{fig:sdtw_roofline_upmem}-right and observe that sDTW is compute-bound in UPMEM. This observation can be attributed to the low-power general-purpose cores in UPMEM that offer poor throughput (146 GINTOPS in contrast to 15700 GINTOPS for the GPU). As arithmetic operations are at the core of sDTW, PNM cannot provide high performance for it. We also observe that UPMEM reduces the energy consumption by 37\% with respect to the GPU by reducing the data movement overheads (\cref{sec:matsa_comparison}). However, poor performance in contrast to the GPU inhibits the effective usability of the platform for the sDTW kernel. 

\vspace{1mm}
\noindent
\setlength{\fboxsep}{7pt}
\setlength{\fboxrule}{1pt}
\fcolorbox{black}{grey}{%
    \parbox{0.93\linewidth}{%
    \noindent\textbf{Key Observation 2:} General-purpose PNM substrates provide higher energy efficiency compared to \mbox{CPU}/\mbox{GPU}/FPGA platforms. However, they fail to offer a high performance solution because of 
the limited arithmetic computation throughput supported by the hardware.
    }       
}
\vspace{1mm}

\subsection{Overcoming bottlenecks in TSA}
\textbf{Need for Processing-Using-Memory (PUM)}. We observe that when executing the sDTW kernel, 1) CPU, GPU, and FPGA platforms are memory-bound, and 2) PNM platforms are compute-bound. In contrast to these platforms, PUM accelerators execute operations directly using the memory cells where data resides~\cite{seshadri2017ambit}. PUM enables 1) exploiting large internal memory bandwidth for memory-bound kernels, and 2) exploiting massive computation parallelism (as high as each bitline) for compute-bound kernels, overcoming key restrictions of CPU, GPU, FPGA and PNM architectures. Based on these observations, \textit{we argue that an accelerator based on PUM is needed to improve TSA's performance and energy efficiency providing a balanced solution}.

\textbf{Cell Technology Choice}. A PUM-based accelerator's performance, energy efficiency, and endurance depend on the underlying substrate's cell technology; thus, it is a critical design choice. Non-Volatile-Memories (NVM) offer a low-energy substrate for PUM as they do not require periodic refresh operations in contrast to DRAM-based PUM~\cite{hajinazar2021simdram, liu2012raidr}. However, it is challenging to support frequent write operations as NVM-based PUM architectures due to significant write latency and low endurance~\cite{zuo2018improving}. Table~\ref{tab:technology_summary} presents the characteristics of NVM technologies we considered for accelerating the sDTW kernel. We discard NAND Flash, ReRAM, and PCM in the first step due to their low endurance and high write latency. Next, we consider FRAM due to its high endurance but discard it due to the high read latency. We then consider MRAM technologies (\cref{sec:mrambackground}) and discard STT-MRAM due to a high write latency. In contrast to STT-MRAM, SOT-MRAM offers 1) high endurance, 2) low read and write latencies, and {3) CMOS compatibility that eases manufacturability~\cite{grossi2016automated}. Considering these characteristics, we argue that SOT-MRAM is a promising substrate for implementing PUM accelerators for kernels with frequent write operations, and evaluate its feasibility for accelerating the sDTW kernel.

\begin{table}[h!]
\centering
\resizebox{\linewidth}{!}{%
\begin{tabular}{|l|c|c|c|}
\hline
    \textbf{Technology} & \begin{tabular}{@{}c@{}}\textbf{Write/Read Energy} \end{tabular} & 
    \begin{tabular}{@{}c@{}}\textbf{Write/Read Time}\end{tabular}
    & 
    \begin{tabular}{@{}c@{}}\textbf{Write Cycles}\end{tabular}\\
\hline\hline
   NAND Flash    &  470pJ / 46pJ    & 200$\mu$s / 25.2$\mu$s  & $10^{5}$  \\\hline
   ReRAM  & 1.1nJ / 525fJ & 10$\mu$s / 5ns   & $10^{5}$  \\\hline
   PCM  & 13.5pJ / 2pJ & 150ns / 48ns   & $10^{7}$  \\\hline
   FRAM  &  1.4nJ / 1.4nJ   & 120ns / 120ns   & $10^{15}$ \\\hline   
   STT-MRAM~  &  2nJ / 34pJ    & 250ns / 10ns   & $>10^{15}$  \\\hline
   \textbf{SOT-MRAM} & 334pJ / 247pJ     & \textbf{1.4ns / 1.1ns}   & $>10^{15}$  \\\hline
\end{tabular}
}
\caption{Characteristics of different NVM technologies~\cite{daulby2020comparing}.}
\label{tab:technology_summary}
\end{table}

We conclude that the MRAM-PUM acceleration approach has the potential to overcome TSA's bottlenecks and provide a faster and more efficient solution than the state-of-the-art.

\subsection{MRAM-based PUM Computation}
\label{sec:mrambackground}
Many prior works demonstrate significant performance and energy efficiency improvements for machine learning workloads via PUM in resistive crossbars~\cite{mittal2018survey} by exploiting matrix-vector multiplication. Other approaches can exploit bitwise operations with high performance and energy savings~\cite{zhang2021time, jin2022high, wang2022reconfigurable}. Figure~\ref{fig:sot_crossbar}-a shows a typical crossbar organization with memory cells connected using bitlines and wordlines.

\begin{figure}[h!]
    \centering
    \includegraphics[width=\linewidth]{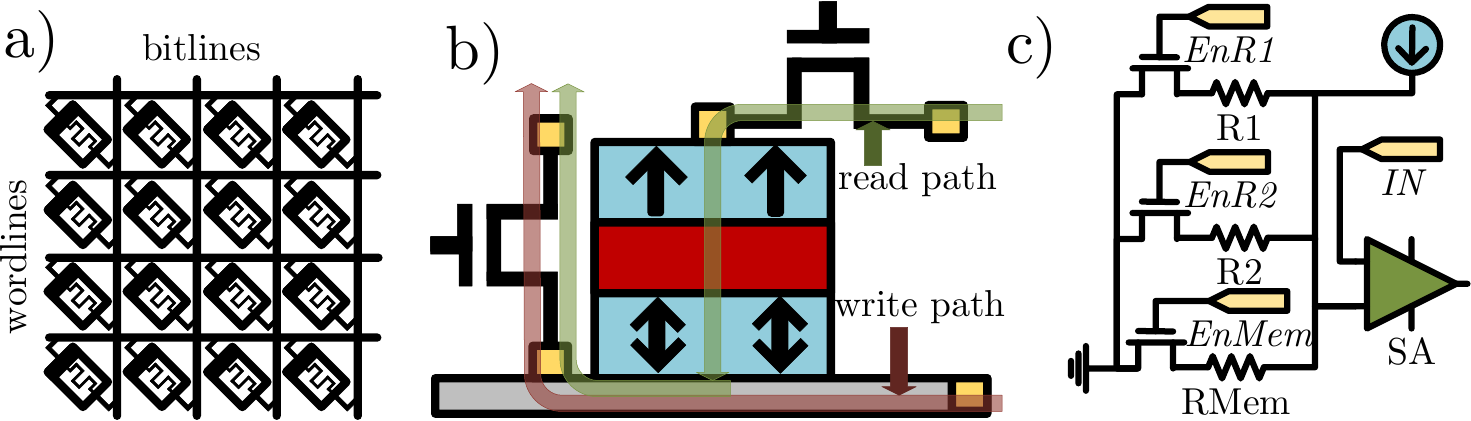}
    \caption{a) Crossbar organization. b) Magneto-resistive cell. c) Reconfigurable SA that performs in-memory operations based on the voltage variations across the bitline.}
    \label{fig:sot_crossbar}
\end{figure}

Figure~\ref{fig:sot_crossbar}-b shows the basic structure of a Spin-Orbit Torque (SOT)-MRAM cell composed of a stack of Magnetic Tunnel Junctions (MTJs) (cyan and red blocks in the figure) and a Heavy Metal Layer (grey block in the figure).

\begin{itemize}
\item  \textbf{Magnetic Tunnel Junction (MTJ)}. Consists of a fixed layer with a pinned magnetization direction, a free layer whose magnetization can be changed, and an insulating tunnel barrier between them.

\item \textbf{Heavy Metal Layer}. This layer is placed next to the MJT to facilitate the spin-orbit torque effect. Common heavy metals used include tantalum (Ta) and tungsten (W).
    
\end{itemize}

 The change of orientation of one of the layers of the stack results in a variation in the device's electrical resistance. However, compared to Spin-Transfer-Torque MTJ (STT-MTJ)~\cite{daulby2020comparing}, SOT-MJT features separated read and write paths, enhancing endurance and widening the read/write margin. Then, sense amplifiers interpret the resulting voltage as boolean:

\begin{itemize}
    \item \textbf{Read Operation}. During a read operation, the resistance of the MTJ is measured. The resistance is sensitive to the relative alignment of the magnetization in the fixed and free layers, allowing the stored data (Boolean values representing 0 or 1) to be read.
    \item \textbf{Write Operation}. During a write operation, an electric current is applied through the heavy metal layer, inducing a spin current. This spin current exerts torque on the free layer, causing its magnetization direction to switch and changing the stored Boolean data.
\end{itemize}

Unlike STT-MTJ, which faces read disturbance issues limiting the read circuit frequency, SOT-MTJ allows for flexible adjustment of current magnitude in the read circuit without concerns about read disturbance effects. As a consequence, it enables more accurate sensing which is crucial to implement in-memory operations. This suggests SOT-MRAM as a better candidate for PUM applications.

\textbf{Bitwise PUM Mechanism.}
The matrix-vector PUM mapping proposed in prior works cannot be applied to dynamic programming (DP) algorithms (e.g., sDTW) since they perform matrix-vector multiplication. DP requires computing a 2D scoring matrix by traversing it row-by-row. Moreover, prior crossbar substrates offer limited support for other operations (e.g., minimum calculation). To overcome this challenge, MAGIC~\cite{kvatinsky2014magic} proposes decomposing complex operations into simple Boolean functions (e.g., AND, NOR, XOR) to support them in the substrate. The key idea is to vertically map the operands (e.g., 32-bit integers) to the crossbars' columns using (typically) one bit per cell (e.g., each operand value takes 32 bits of a given column). Then, the desired operation (e.g., addition) is decomposed to simple bitwise operations (e.g., NOR) and performed bit-by-bit via sequentially activating two cells for each operand simultaneously. This approach creates a difference in the voltage over the bitline depending on the content of the activated cells, which depends on the resistance they hold. Then, a modified sense amplifier calculates the result based on that voltage difference and thresholds, storing it in a cell of the same column. While this process is inherently sequential and the latency per operation is higher than a CMOS-based approach, the 1) independence across columns and 2) the lack of data movement enables immense parallelism and, thus, an overall higher throughput than CMOS-based solutions. Figure~\ref{fig:sot_crossbar}-c shows a sense amplifier (SA) slightly modified with respect to commodity ones, including different voltage thresholds for the operations.
\section{\myName{} Architecture}
\subsection{Overview}
\begin{figure*}[t!]
    \centering
    \includegraphics[width=\textwidth]{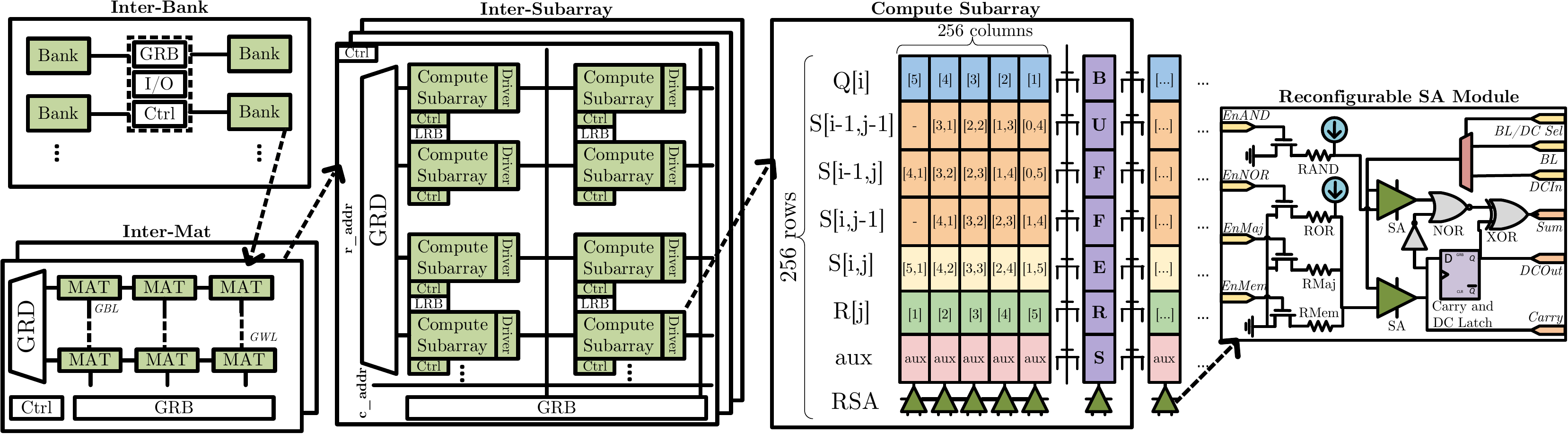}
    \caption{\myName{}'s high-level architecture and data mapping flow. 
    }
    \label{fig:matsa_architecture}
\end{figure*}

\myName{} is an \underline{M}RAM-based \underline{A}ccelerator for \underline{T}ime \underline{S}eries \underline{A}nalysis. Figure~\ref{fig:matsa_architecture} presents an overview of our proposed architecture. \myName{} is composed of several chips divided into multiple \textit{banks}. Banks belonging to the same chip share buffers and I/O interfaces and work in a lock-step approach. Each bank is composed of several \textit{Multiple Memory Matrices (MATs)}. The MATs share a \textit{Global Row Buffer (GRB)} and are connected to a \textit{Global Row Decoder (GRD)}. We place a \textit{Local Row Buffer (LRB)} for every pair of subarrays to improve performance. Each subarray is composed of magnetoresistive devices that are connected to the \textit{Write Word Lines} (WWL), \textit{Write Bit Lines} (WBL), \textit{Read Word Lines} (RWL), \textit{Read Bit Lines} (RBL), and \textit{Source Lines} (SL). The compute-enabled subarrays perform the sDTW computation using Reconfigurable Sense Amplifiers (RSAs).

The execution flow is orchestrated by a hierarchy of small controllers implemented as finite state machines (FSMs). \myName{} comprises of 1) a global controller
that orchestrates inter-bank flow, 2) inter-mat controllers that take care of the inter-mat flow, and 3) subarray controllers that activate the memory rows and
drive the RSAs to run sDTW’s algorithm.

\subsection{\myName{} Subarrays}
\myName{} subarrays are comprised of MRAM cells following a crossbar organization and can work either in regular memory or compute mode. This is a desirable feature since our design consists of 1) subarrays that temporarily buffer the data until they are being processed and 2) subarrays that perform the actual computation. Adjacent subarrays are connected using pass gates and aux columns (purple one in Figure~\ref{fig:matsa_architecture}) to enable the data flow through the hierarchy.

\textbf{Memory Subarrays}. \myName{} subarrays in \textit{regular memory mode} support both read and write data operations and work in the same way as conventional non-PUM-enabled memory.

\textbf{Compute Subarrays}. \myName{} subarrays working in \textit{compute mode} perform bit-wise operations on input data located in cells of the same column. This enables the parallel execution of many operations since all columns in the subarray work in parallel. The key idea is to select two or three input values simultaneously using the Memory Row Decoder (MRD). This produces an equivalent resistance that depends on the content of the selected cells and modifies the sensing voltage across the column accordingly. \myName{}'s Ctrl can select different operations from the Reconfigurable Sense Amplifiers (RSAs) that are placed per column. We modify the RSAs to execute operations by equipping them with different resistances to model the voltage thresholds, logic gates (i.e., NOR, XOR, INV), a register, and a multiplexer (see Figure~\ref{fig:matsa_architecture}). The RSAs in Compute subarrays support the same operations as memory subarray RSAs, enabling switching between operating in compute and memory modes.

\subsection{PUM Operations}
\label{sect:operations}
\myName{} implements the following PUM operations to support the execution of sDTW  (detailed in Algorithm~\ref{alg:sdtw}):
\begin{itemize}
    \item \textbf{Vertical Row Copy}. MATSA executes consecutive memory read and write operations in the same cycle to improve performance by activating two rows simultaneously. In the first half cycle, the subarray's MRD activates the source row read by the LRB. Next, the destination row is activated to store the data in the second half cycle. This mechanism works at MAT and bank levels using the Global Row Buffer (GRB) to accelerate the copies across the hierarchy.

    \item \textbf{Diagonal Row Copy}. The Ctrl executes a diagonal copy shift data between adjacent columns. The Ctrl leverages the available registers in the RSA and the interconnections between the RSAs. The operation is executed in two steps. First, the RSA reads the value in the source column. Second, the destination RSA (in an adjacent column) reads the value from the source RSA and writes it to its column.

    \item \textbf{Addition/Subtraction}. \myName{} executes Bit-serial addition/subtraction across columns. The Ctrl executes operations starting from the least significant bit of the two operands until the most significant bit. Every bit operation requires two memory cycles, further divided into four half cycles. In the first half-cycle, the RSAs read voltage difference across all cells activated in the same bit lines as input operands and calculate the \textit{Sum}. The RSA updates the Sum based on the stored Carry value in the register. In the second half-cycle, the RSAs write the \textit{Sum} value to the destination cell. In the third half-cycle, the RSAs calculate the new \textit{Carry} value based on a majority function of the operand rows and an auxiliary row reserved for the Carry bit. In the fourth half-cycle, RSAs write the new Carry value in the auxiliary row for the next Carry calculation.
    
    \item \textbf{Absolute Calculation}. To calculate the absolute value, \myName{} first checks the sign bit, leading to two possible scenarios: 1) if the number is positive, no change is needed; otherwise, 2) if the number is negative, \myName{} inverts the bits of the number and adds '1' to the result (similar to 2's complement).
    
    \item \textbf{Minimum Value}. To calculate the minimum value between three elements, \myName{} performs two comparisons based on the subtraction operation. First, it calculates the difference between the two numbers. Second, it checks the resulting sign from the previous step and selects one of the two numbers for comparison against the third. The final comparison sign determines the minimum between three values. The logic can be similarly extended for comparing more than three values.
\end{itemize}

\subsection{Data Mapping}
Section~\ref{sec:bottlenecks} demonstrates that sDTW is an embarrassingly parallel algorithm. We design \myName{}'s data mapping to leverage MRAM's parallel column-wise computation capability. Three data structures are involved in the sDTW computation: 1) reference sequence (of length $O$(M)), 2) query sequence (of length $O$(N)), and 3) the warping matrix (dynamic programming matric with size $O$(NM)). The data structures are mapped to the subarray as follows:

\begin{itemize}
    \item \textbf{Reference Elements (\texttt{R[j]})}. We vertically map each reference element to 32 cells of a column. If 1) the number of available columns is larger than the number of elements in reference, we replicate the reference to multiple columns to increase parallelism (distributing the queries between them). If 2) the number of available columns is lower than the number of elements in reference, we divide the query and complete the process in sequential batches. No action is needed if 3) available columns are equal to the number of elements in reference.
    
    \item \textbf{Query Elements (\texttt{Q[i]})}. We vertically map each query element to 32 cells of a column. New query elements are introduced on the left side of the crossbar, and they are right-shifted in each successive step (see \cref{sec:exec_flow}).

    \item \textbf{Current S\_vector (\texttt{S[i, j])}}. We define the current vector of the warping matrix as the S\_vector. We vertically map each element of the S\_vector to 32 cells of a column, being aligned with the query and reference elements (i and j indexes, respectively). 
    
    \item \textbf{Temporal S\_vectors (\texttt{S[i-1, j-1], S[i-1, j], S[i, j-1]})}. We vertically map the three temporal vectors along the reference and query elements. Mapping the temporal vectors in the same subarray leverages parallelism in the subarray as each column can compute lines 8-9 of Algorithm~\ref{alg:sdtw} completely in parallel. Then, those vectors are efficiently updated also in parallel for the next iteration of the loop thanks to the vertical and diagonal row copies.
    
    \item \textbf{Aux Cells}. Each column has a slice of 64 cells used to hold the partial results during the execution flow.
\end{itemize}

We calculate the distance between each data point in the reference and the query by iterating over the current S\_vector of the warping matrix (see Algorithm~\ref{alg:sdtw}). Each element in the S\_vector (mapped across different crossbar columns) requires accessing previous S\_vector  values that are mapped to the same column (i.e., $S[i-1,j]$) and adjacent columns (i.e., $S[i,j-1]$, $S[i-1,j-1]$). To break this data dependency, we add three temporal S\_vectors in the crossbar array that are updated in each step of the computation: $S[i-1,j-1]$, $S[i-1,j]$ and $S[i,j-i]$ (see Figure~\ref{fig:matsa_architecture}). Overall, our optimization reduces the memory footprint from $O$(NM) (whole matrix) to $O$(4M) (S\_vector plus three aux ones). 

\subsection{Execution Flow}
\label{sec:exec_flow}
\myName{}'s execution flow follows a wavefront approach~\cite{dios2012case}, which reflects the computation pattern in dynamic programming applications. The motivation is that sDTW's matrix has to be computed in the wavefront manner due to inter-cell dependencies. Figure~\ref{fig:wavefront} shows an example of how we tackle this restriction by assuming one reference time series (red one) and two queries (green and ocher).
\begin{figure}[h!]
\centering
\includegraphics[width=\linewidth]{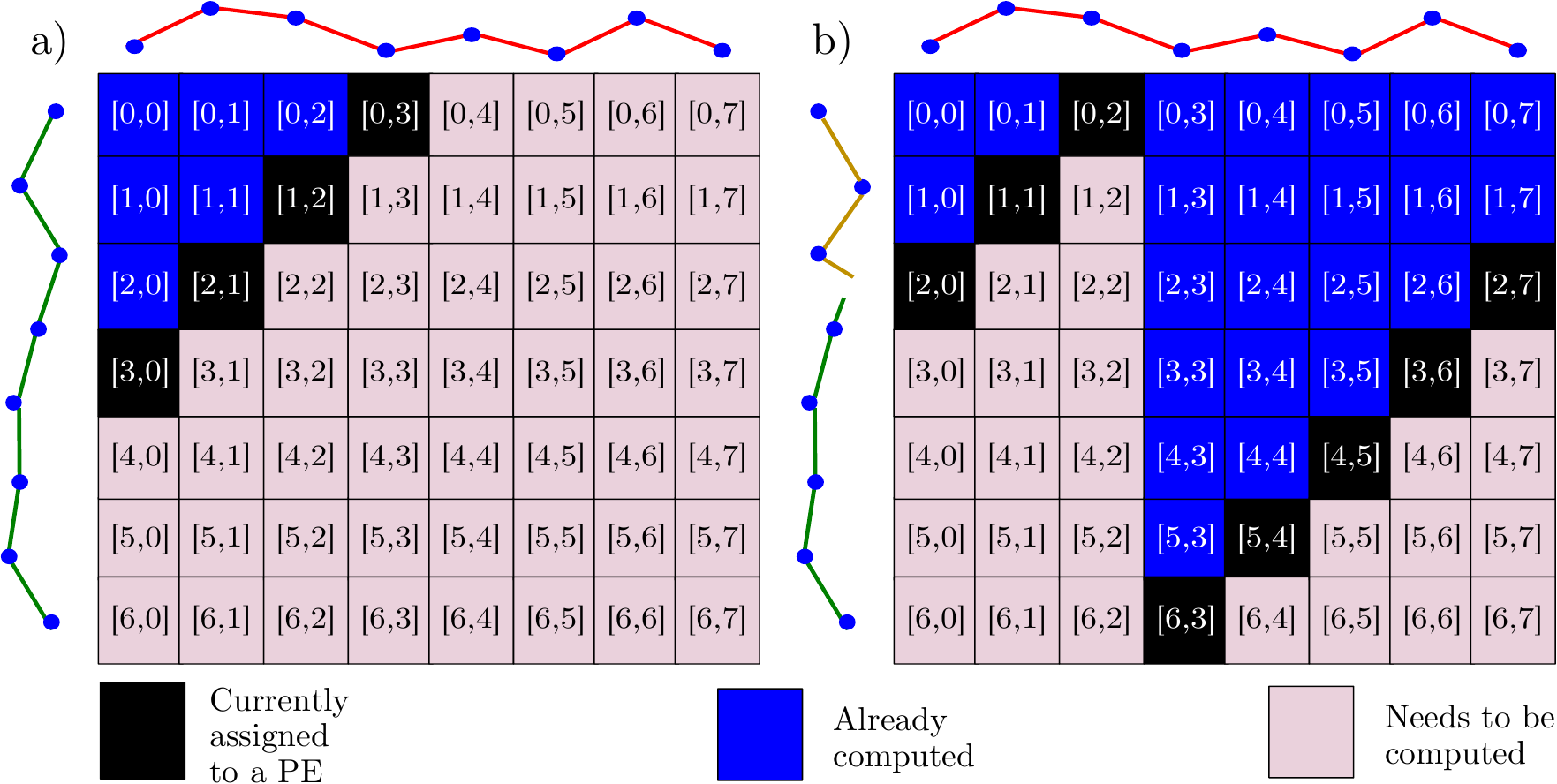}
\caption{Wavefront-based sDTW computation in \myName{}.
}
\label{fig:wavefront}
\end{figure}

The key idea is to make computation flow diagonally by assigning one element in the wavefront to each processing element (PE), and using the \textit{diagonal row copy} operation (\cref{sect:operations}) to shift data between columns on the wavefront. This is needed since each cell requires taking values from its left column, thus their data values need to be available prior to computation. Because of that, each PE advances computation in the vertical direction with one cell delay with its left PE, ensuring that the data needed to calculate the next value is available. Figure~\ref{fig:wavefront}-a shows an initial state where the computation just started. In this example, only PEs where their column contain black rectangles are are performing computation. Note that in every step the wavefront introduces a new PE to the active set, achieving maximum performance after \textit{number\_of\_PEs} steps. When reaching point, all PEs are able to perform useful work in a given execution step. \mbox{Figure~\ref{fig:wavefront}-b} shows how this initialization phase can be amortized by pipelining. By introducing a new query to compare against the reference before the prior one finishes, \myName{} ensures that all PEs have work to do even during the transitions between queries. Overall, this execution flow enables 1) leveraging the subarray columns in parallel for the query, and 2) creation of an inter-subarray pipeline to leverage parallelism across queries, i.e., by processing queries in parallel. The execution flow of each cell goes through the following steps:

\begin{enumerate}

    \item \textbf{Distance Calculation}. Calculation of $dist(Q[i], R[j])$, which provides the first partial result $P1$. This process implies several substeps depending on the selected distance metric, (e.g., subtraction $\rightarrow$ absolute value).
    
    \item \textbf{Minimum}. Calculation without storing the result of $min(S[i-1,j-1], S[i-1,j], S[i,j-1])$, which produces the value for the next step $S1$.
    
    \item \textbf{Addition}. Calculation of the addition between the minimum value selected in the previous step ($S1$) and the partial result $P1$. 
    
    \item \textbf{Diagonal Copy}. Copying the $S[i,j]$ vector into the $S[i,j-1]$ vector shifted by one to the right.
    
    \item \textbf{Diagonal Copy}. Copying the $S[i-1,j]$ vector into the $S[i-1,j-1]$ vector shifted by one to the right.
    
    \item \textbf{Vertical Copy}. Copying the $S[i,j]$ vector into the $S[i-1,j]$ vector.
    
    \item \textbf{Diagonal Copy}.
    Copying the $Q[i]$ vector into the same $Q[i]$ vector but shifted one position to the right.
\end{enumerate}

\subsection{Programming Interface \& System Integration}
\textbf{Programming Interface}. We expose an API (Listing~\ref{list:api}) that allows to invoke \myName{} from the host processing unit.

\begin{lstlisting}[caption={\myName{}'s host interface function.},captionpos=b,label={list:api}]
void matsa(DTYPE * ref, DTYPE * queries,  uint64_t * ref_size, uint64_t * query_sizes, uint64_t n_queries, char * mode, char * dist_metric, DTYPE anomaly_thres, bool * anomalies, DTYPE * distances)
\end{lstlisting}

\myName{} expects input data to be in a supported type/precision DTYPE (integer: \texttt{int8}, \texttt{int16}, \texttt{int32} or \texttt{int64}; fixed-point: \texttt{fp32} or \texttt{fp64}), the selected mode (either \textit{query\_filtering}, where queries are compared against the reference or \textit{self\_join}, where slices of the reference are compared against themselves) and the distance metric (\textit{abs\_diff} or \textit{square\_diff}). \myName{} can also take an anomaly threshold, which returns an array with the detected ones.

\textbf{System Integration}. \myName{} is designed to work synergistically with the CPU to accelerate TSA. We propose three  \myName{} versions to meet the requirements of different environments, as we describe next.
\begin{enumerate}[label=\alph*)]
\item \textbf{\myName{}-HPC}. A high-performance PCIe-based accelerator intended to be integrated into servers.
\item \textbf{\myName{}-Embedded}. A small chip intended to be integrated with edge devices (e.g., sensors).
\item \textbf{\myName{}-Portable}. A USB-based accelerator intended for use in desktops and laptop computers. 
\end{enumerate}
\section{Evaluation}
\label{sec:evaluation}
\subsection{Methodology}
To comprehensively quantify the performance and energy efficiency improvements of MATSA, we compare it with the following systems.
\begin{itemize}
    \item \textbf{CPU-ARM (\texttt{cpuarm}):} 4-core ARM CPU @ 2.5GHz, 32KB L1 and 8GB LPDDR4.
    
    \item \textbf{CPU-i7 (\texttt{cpui7}):} 6-core (12 threads) Intel i7 x86 CPU @ 3.2GHz, 64KB L1, 256KB L2, 12MB L3 and 64GB DDR4.
    
    \item \textbf{CPU-Xeon (\texttt{cpuxeon}):} Two 18-core (36 threads) Intel Xeon Gold 6154 x86 CPUs @ 3GHz, 32KB L1, 1MB L2, 24.75 MB L3 and 768GB DDR4.

    \item \textbf{GPU (\texttt{gpu}}): NVIDIA Tesla V100 with 32GB of HBM.
 
    \item \textbf{FPGA (\texttt{fpga}}): Xilinx Alveo U50 with 8GB HBM memory.

    \item \textbf{UPMEM (\texttt{upmem}):} Server-class Processing-Near-Memory DIMMs with 2560 DPUs running at 425MHz~\cite{devaux2019true}. 
    
    \item \textbf{\myName{}-Embedded (\texttt{matsa-embedded})}: consisting of 128 compute-enabled crossbars (1MB) and 896 regular-memory crossbars (7MB).

    \item \textbf{\myName{}-Portable (\texttt{matsa-portable})}: consisting of 1024 compute-enabled crossbars (8MB) and 7168 regular-memory crossbars (56MB).
    
    \item \textbf{\myName{}-HPC (\texttt{matsa-hpc})}: consisting of 4096 compute-enabled crossbars (32MB) and 28672 regular-memory crossbars (224MB).
\end{itemize}

\textbf{Baselines}. We use ZSim+Ramulator~\cite{zsim+ramulator} and McPAT for the \texttt{cpuarm} platform. For the \texttt{cpui7} and \texttt{cpuxeon} platforms, we have access to the target hardware and measure performance and energy consumption values by averaging five repeated executions. The energy consumption is determined using Intel RAPL tools. To evaluate the performance of the \texttt{upmem} platform, we implement and optimize the sDTW algorithm as shown in Algorithm~\ref{alg:sdtw}. To evaluate the performance on the \texttt{fgpa} platform, we implement the sDTW algorithm using High-Level Synthesis vendor tools from Xilinx and optimize the implementation to utilize eight compute units and maximize the utilization of the available HBM bandwidth. We evaluate the performance of the \texttt{gpu} platform by optimizing a CUDA-based implementation of sDTW to maximize the HBM bandwidth utilization via memory coalescing. We measure the GPU's energy consumption using the \textit{NVIDIA-smi} tool.

\textbf{\myName{}}. Due to the lack of a cycle-accurate simulator for MRAM-based accelerators, we implement an in-house simulator for MRAM-based PUM. Figure~\ref{fig:matsasim} shows an overview this simulator. We provide the workload characteristics and the MRAM device characteristics under study, and the simulator computes the performance and energy efficiency in return. We plan to release this simulator for public use of the community after acceptance of this work.

We perform a sensitivity analysis by sweeping MRAM devices' latency and energy from conservative to optimistic values based on MRAM device trends~\cite{saha2022comparative} listed in Table~\ref{tab:matsa_parameters}.  Based on that, we conservatively select an operating point (highlighted in bold) for the evaluations taking into account realistic MRAM device progress projections~\cite{endoh2020recent}. We input the workload parameters and MRAM characteristics obtained from the parameter sweep to the simulator to get the workload's execution time and energy consumption.

\begin{figure}[h!]
    \centering
    \includegraphics[width=\linewidth]{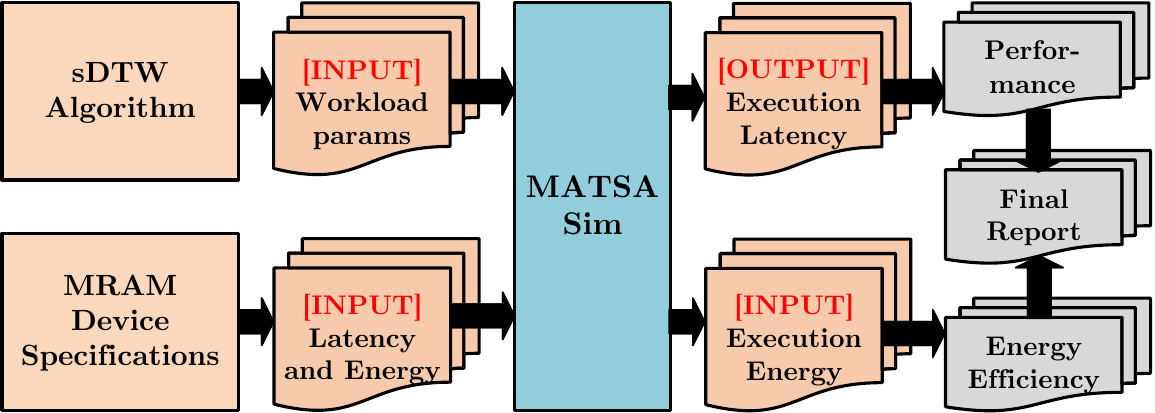}
    \caption{Overview of MATSA Simulator.}
    \label{fig:matsasim}
\end{figure}

\begin{table}[h!]
\centering
\resizebox{0.85\linewidth}{!}{
\begin{tabular}{|l|c|}
\hline
    \textbf{Parameter} & \textbf{Values}\\
\hline\hline
    Crossbar Size (cells)       &  256x256\\
    \hline
    Number of Crossbars  &  128, 256, 512, 1024, 2048, 4096\\
    \hline
    Read Latency (ns)    &  1, 3, \textbf{5}, 10, 20\\
    \hline
    Write Latency (ns)   &  1, 3, 5, \textbf{10}, 20\\
    \hline
    Read Energy (pJ)     &  20, \textbf{50}, 100\\
    \hline
    Write Energy (pJ)    &  30, \textbf{70}, 400\\
\hline
\end{tabular}
}
\caption{\myName{} design space exploration parameters.}
\label{tab:matsa_parameters}
\end{table}

\noindent\textbf{Datasets}. We perform \myName{}'s design exploration using the datasets in Table~\ref{tab:workloads}, which ease understanding of the tradeoffs. Then, we compare \myName{} against baselines in real scenarios using the real datasets in Table~\ref{tab:workloads_real}. The data type for these evaluations is \texttt{int32}, which covers the data ranges of all the evaluated the workloads.

\begin{table}[h!]
\centering
\resizebox{0.75\linewidth}{!}{
\begin{tabular}{|l|c|}
\hline
    \textbf{Parameter} & \textbf{Values}\\
\hline\hline
    Reference Size     &  64K, 128K, 256K, 512K\\
    \hline
    Query Size         &   4K, 8K, 16K, 32K\\
    \hline
    Number of Queries  &   4K, 8K, 16K, 64K\\
\hline
\end{tabular}
}
\caption{Workloads used in \myName{} characterization.}.
\label{tab:workloads}
\end{table}

\begin{table}[h!]
\centering
\resizebox{\linewidth}{!}{%
\begin{tabular}{|l|c|c|c|}
\hline
    \textbf{Time Series} & \textbf{Reference Size}& \textbf{Query Size}& \textbf{Num.  Queries}\\
\hline\hline
   Human~\cite{veeraraghavan2006function}    &  7997    & 120  & 128K \\\hline
   Song~\cite{YHK16}           &  20234   & 200  & 64K \\\hline
   Penguin~\cite{PenguinData}  &  109842  & 800  & 32K \\\hline
   Seismology~\cite{YHK16}        &  1727990 & 64   & 16K \\\hline
   Power~\cite{murray2015data}           &  1754985 & 1536 & 16K \\\hline
   ECG~\cite{TDE+92}               &  1800000 & 512  & 16K \\
\hline
\end{tabular}
}
\caption{Real-world workloads used in our evaluation.}
\label{tab:workloads_real}
\end{table}

\subsection{\myName{} Characterization}
\label{subsect:matsa_charact}
We perform a design space exploration of \myName{} taking into consideration performance parameters of the cells (i.e., read/write latencies and energies). 

\textbf{Read/Write Latencies.} We evaluate how changing the read/write latencies affects the execution time and present the results in Figure~\ref{fig:exec_latencies}. We observe that, increasing read latency by 10$\times$ incurs a 4.7$\times$ execution time penalty, while increasing the
write latency incurs a 6.5$\times$ penalty.

\begin{figure}[h!]
    \centering
    \includegraphics[width=\linewidth]{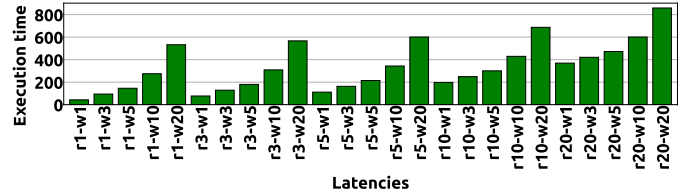}
    \caption{Execution time when varying cell read and write latencies (ref\_size=128K, query\_size=8K, num\_queries=8K, matsa\_cols=128K).}
    \label{fig:exec_latencies}
\end{figure}

\vspace{1mm}
\noindent
\setlength{\fboxsep}{7pt}
\setlength{\fboxrule}{1pt}
\fcolorbox{black}{grey}{%
    \parbox{0.93\linewidth}{%
\noindent\textbf{Key Observation 3:} using a low write latency memory technology is crucial for \myName{}'s design.
    }       
}
\vspace{1mm}

\textbf{Read/Write Energies.} We evaluate how the total execution energy varies with the per word write/read energy, and show the results in Figure~\ref{fig:exec_energies}. We observe here that the contributions of read energy and write energy are similar, thus both of them have to be carefully taken into consideration.

\begin{figure}[h!]
    \centering
    \includegraphics[width=\linewidth]{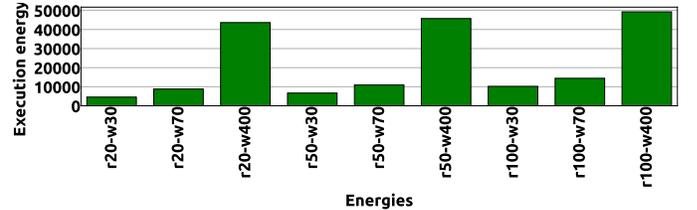}
    \caption{Execution energy when varying cell read and write energies (ref\_size=128K, query\_size=8K, num\_queries=8K, matsa\_cols=128K).}
    \label{fig:exec_energies}
\end{figure}

\vspace{1mm}
\noindent
\setlength{\fboxsep}{7pt}
\setlength{\fboxrule}{1pt}
\fcolorbox{black}{grey}{%
    \parbox{0.93\linewidth}{%
\noindent\textbf{Key Observation 4:} read energy contributes 45\% and write energy contributes 55\% to the total energy consumption of a given execution.
    }       
}
\vspace{1mm}

\textbf{Dataset Sizes.} First, we evaluate how the execution time varies with different dataset sizes (i.e., ref\_size and query\_size) and present the results in Figure~\ref{fig:exec_workload_sizes}. Second, we evaluate how the execution energy varies with different dataset sizes and present the results in Figure~\ref{fig:energy_workload_sizes}. We observe that both reference size and  query size contribute equally to the execution time and energy. This happens because the total number of operations needed is directly proportional to ref\_size$\times$query\_size. Our observation corroborates our earlier analysis stating that query-specific sDTW implementations do not fairly represent GPU performance, and there is a need for a more general solution.

\begin{figure}[h!]
    \centering
    \includegraphics[width=\linewidth]{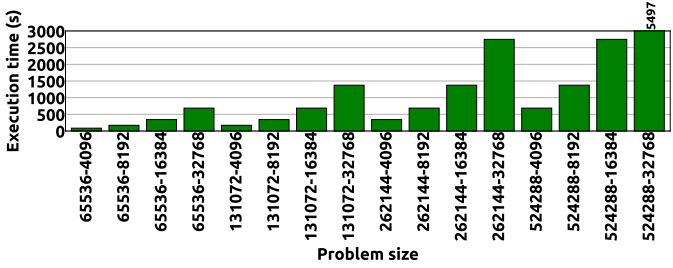}
    \caption{Execution time when varying dataset sizes (num\_queries=8K, matsa\_cols=128K).}
    \label{fig:exec_workload_sizes}
\end{figure}

\begin{figure}[h!]
    \centering
    \includegraphics[width=\linewidth]{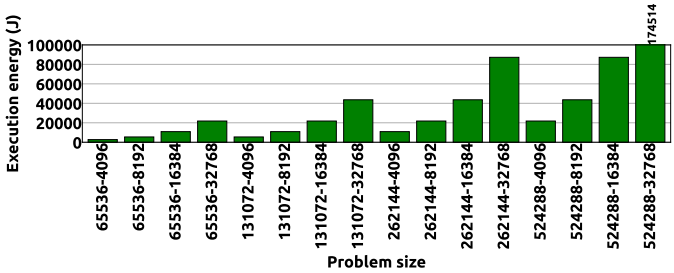}
    \caption{Execution energy when varying dataset sizes (num\_queries=8K, matsa\_cols=128K).}
    \label{fig:energy_workload_sizes}
\end{figure}

\vspace{1mm}
\noindent
\setlength{\fboxsep}{7pt}
\setlength{\fboxrule}{1pt}
\fcolorbox{black}{grey}{%
    \parbox{0.93\linewidth}{%
\noindent\textbf{Key Observation 5:} Total execution time and energy consumption are proportional to both ref\_size and the query\_size.}
}
\vspace{1mm}

\textbf{\myName{} sizes.} We evaluate how the execution time varies when changing the number of \myName{}'s compute-enabled columns in Figure~\ref{fig:exec_matsa_sizes}. \myName{} provides almost-ideal scaling.

\begin{figure}[h!]
    \centering
    \includegraphics[width=\linewidth]{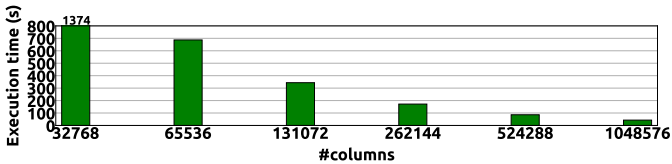}
    \caption{Execution time when varying \myName{} sizes.}
    \label{fig:exec_matsa_sizes}
\end{figure}

\vspace{1mm}
\noindent
\setlength{\fboxsep}{7pt}
\setlength{\fboxrule}{1pt}
\fcolorbox{black}{grey}{%
    \parbox{0.93\linewidth}{%
    \noindent\textbf{Key Observation 6:} Bit-serial computation across columns enables almost-ideal scaling when increasing the size of the workload.
    }       
}
\vspace{1mm}

\textbf{Endurance.} Assuming that \myName{} is built using 5/10ns rd/wr cells and runs 24/7 for ten years, we estimate that each cell will be written $\approx 4 \times 10^{9}$ times. Based on Table~\ref{tab:technology_summary}, limited-endurance cells (e.g., ReRAM) would fail within one day. In contrast, high-endurance cells ($10^{15}$ writes for SOT-MRAM) can provide a very large usable lifetime.

\textbf{Hardware Overheads.} MATSA introduces hardware overheads in two components: 1)  Reconfigurable SAs and 2) MATSA controllers. Reconfigurable SAs add 13 transistors to a traditional SA, thus taking into consideration typical SA and cell areas~\cite{uddin2018practical,seo2016area}, our design increases the overall crossbar area by less than 1\%. MATSA controllers are implemented as small finite-state machines whose area is negligible compared to the memory arrays.

\subsection{System Evaluation}
\label{sec:matsa_comparison}

\textbf{\myName{}-Embedded and \myName{}-Portable}. We compare the performance of \myName{}-Embedded (32K compute-enabled columns) and \myName{}-Portable (256K compute-enabled columns) with \texttt{cpuarm}, \texttt{cpui7}, and \texttt{fpga} baselines in Figure~\ref{fig:matsa_emb_port}a. The smallest version, \myName{}-Embedded, provides 30.20$\times$/1.30$\times$/8.14$\times$ lower execution times than \texttt{cpuarm}, \texttt{cpui7}, and \texttt{fpga}, respectively. \myName{}-Portable is further able to improve the performance by 241.66$\times$/10.40$\times$/65.28$\times$ with respect to the same baselines, respectively. These performance improvements stem from the higher available parallelism in PUM, where all compute-enable columns can compute independently. Next, we compare the energy consumption of \myName{}-Embedded and \myName{}-Portable with the same baselines in Figure~\ref{fig:matsa_emb_port}b. \myName{}-Embedded reduces the energy consumption by 45.67$\times$/10.64$\times$/24.58$\times$ with respect to \texttt{cpuarm}, \texttt{cpui7} and \texttt{fpga} baselines, respectively. We observe that 1) the energy reduction comes from eliminating the expensive off-chip data movement and 2) \myName{}-Portable reduces the energy consumption by roughly the same factor as \myName{}-Embedded. We deduce from these results that scaling \myName{} improves the performance but does not penalize the energy efficiency.

\begin{figure}[h!]
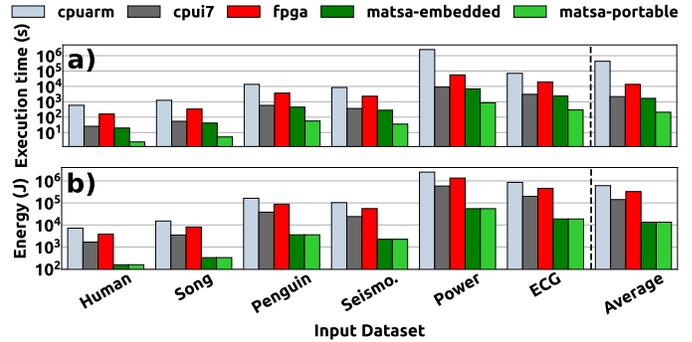

\begin{subfigure}{\linewidth}
  \centering
 \includegraphics[width=\linewidth]{figures/performance_real_matsa-emb-port.pdf}
\end{subfigure}
\begin{subfigure}{\linewidth}
  \centering
  \includegraphics[width=\linewidth]{figures/energy_real_matsa-emb-port.pdf}
  \end{subfigure}
\caption{Latency and energy consumption of \myName{}-Embedded (num\_cols=32K) and \myName{}-Portable (num\_cols=256K) versus baselines (rd\_lat=5ns, wr\_lat=10ns, rd\_en=50nJ, wr\_en=70nJ).}
\label{fig:matsa_emb_port}
\end{figure}

\textbf{\myName{}-HPC}. We first perform a performance comparison of \myName{}-HPC and present the results in Figure~\ref{fig:matsa_hpc}a.
\begin{figure}[h!]
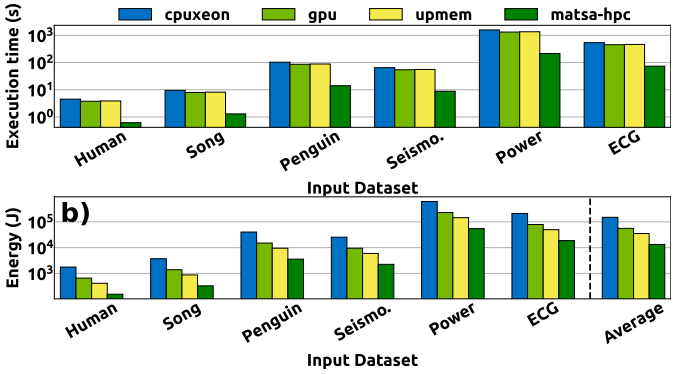

\begin{subfigure}{\linewidth}
  \centering
  \includegraphics[width=\linewidth]{figures/performance_real_matsa-hpc.pdf}
\end{subfigure}
\begin{subfigure}{\linewidth}
  \centering
  \includegraphics[width=\linewidth]{figures/energy_real_matsa-hpc.pdf}
\end{subfigure}
\caption{Execution times and energy consumption of \myName{}-HPC (num\_cols=1M) versus baselines (rd\_lat=5ns, wr\_lat=10ns, rd\_en=50nJ, wr\_en=70nJ).}
\label{fig:matsa_hpc}
\end{figure}
We observe that \myName{}-HPC achieves 7.3$\times$/6.15$\times$/6.3$\times$ lower execution times than \texttt{cpuxeon}, \texttt{gpu} and \texttt{upmem}, respectively, owing to enormous available parallelism (one million compute columns). 
Second, we compare the energy consumption of \myName{}-HPC in Figure~\ref{fig:matsa_hpc}b and observe that it provides 11.29$\times$/4.21$\times$/2.65$\times$ lower energy consumption than \texttt{cpuxeon}, \texttt{gpu} and \texttt{upmem}, respectively. The energy efficiency benefits of \myName{}-HPC stem from the elimination of the off-chip data movements. We note that \texttt{cpuxeon} is bottlenecked by 1) the limited parallelism (number of cores) and 2) the high data movement costs through the memory hierarchy. The \texttt{gpu} baseline provides high parallelism but is limited by data movement from and to memory. The PNM-based \texttt{upmem} baseline provides high parallelism and lowers data access costs compared to CPU and GPUs. However, the sDTW kernel is compute-bound in \texttt{upmem} due to small general-purpose cores, in contrast to MATSA, a dedicated accelerator design for the sDTW kernel.

\textbf{\myName{} Benefits}. Table~\ref{tab:per_energ_summary} summarizes \myName{}'s benefits.

\begin{table}[h!]
\centering
\resizebox{\linewidth}{!}{%
\begin{tabular}{|l|l|c|c|}
\hline
    \textbf{MATSA Version} & \textbf{Baseline}& \textbf{Speedup}& \textbf{Energy Savings}\\
    \hline\hline
   Embedded    &  cpuarm    & 30.20$\times$  & 45.67$\times$  \\\hline
   Portable &  cpui7   & 10.41$\times$   & 10.65$\times$  \\
   &  FPGA   & 65.01$\times$   & 24.58$\times$  \\\hline
   &  Xeon  & 7.35$\times$   & 11.29$\times$  \\
   HPC &  UPMEM  & 6.31$\times$   & 2.65$\times$  \\
   &  GPU  & 6.15$\times$   & 4.21$\times$  \\
  \hline
\end{tabular}
}
\caption{{\myName{}'s Speedup and Energy over baselines.}}
\label{tab:per_energ_summary}
\end{table}
\section{Related Work}
To our knowledge, \myName{} is the first sDTW accelerator via MRAM-based PUM. We compare extensively to CPU, GPU, FPGA, and state-of-the-art PNM platforms in \cref{sec:evaluation}. In this section, we describe related works focusing on accelerating sDTW and prior PUM-based accelerators.

\textbf{Accelerating Dynamic Time Warping (DTW)}. Several works attempt to accelerate the sDTW kernel using GPUs~\cite{SH20, sadasivan2023accelerated} and FPGAs~\cite{9153280, 10.1145/2435264.2435277}. \cref{sec:evaluation} demonstrates that \myName{} improves upon the performance of GPUs and FPGAs by 6.15$\times$ and 65.28$\times$ respectively, and supports arbitrary-sized datasets, a key drawback of prior work. 

\textbf{Processing Near/Using Memory.} 
There has been a significant interest in Processing-[Near/Using]-Memory-based solutions for overcoming the von Neumann bottleneck in modern computation platforms~\cite{mutlu2019processing,ghose2019processing,cali2020genasm,singh2019napel,singh2020nero,stone1970logic, Kautz1969, shaw1981non, kogge1994, gokhale1995processing, patterson1997case, oskin1998active, kang1999flexram, Mai:2000:SMM:339647.339673,murphy2001characterization, Draper:2002:ADP:514191.514197,aga.hpca17,eckert2018neural,fujiki2019duality,kang.icassp14,seshadri2017ambit,seshadri.arxiv16,Seshadri:2015:ANDOR,seshadri2013rowclone,angizi2019graphide,kim.hpca18,kim.hpca19,gao2020computedram,chang.hpca16,xin2020elp2im,li2017drisa,deng.dac2018,hajinazar2021simdram,rezaei2020nom,wang2020figaro,ali2019memory,li.dac16,angizi2018pima,angizi2018cmp,angizi2019dna,levy.microelec14,kvatinsky2014magic,shafiee2016isaac,kvatinsky.iccd11,kvatinsky.tvlsi14,gaillardon2016plim,bhattacharjee2017revamp,hamdioui2015memristor,xie2015fast,hamdioui2017myth,yu2018memristive,giannoula2021syncron,fernandez2020natsa,kim.bmc18,ahn.pei.isca15,ahn.tesseract.isca15,boroumand.asplos18,boroumand2019conda,asghari-moghaddam.micro16,DBLP:conf/sigmod/BabarinsaI15,chi2016prime,farmahini2015nda,gao.pact15,DBLP:conf/hpca/GaoK16,gu.isca16,guo2014wondp,hashemi.isca16,cont-runahead,hsieh.isca16,kim.isca16,kim.sc17,DBLP:conf/IEEEpact/LeeSK15,liu-spaa17,morad.taco15,nai2017graphpim,pattnaik.pact16,pugsley2014ndc,zhang.hpdc14,zhu2013accelerating,DBLP:conf/isca/AkinFH15,gao2017tetris,drumond2017mondrian,dai2018graphh,zhang2018graphp,huang2020heterogeneous,zhuo2019graphq,santos2017operand,ghoseibm2019,wen2017rebooting,besta2021sisa,ferreira2021pluto,olgun2021quactrng,lloyd2015memory,elliott1999computational,zheng2016tcam,landgraf2021combining,rodrigues2016scattergather,lloyd2018dse,lloyd2017keyvalue,gokhale2015rearr,nair2015active,jacob2016compiling,sura2015data,nair2015evolution,balasubramonian2014near,xi2020memory,impica,boroumand2016pim,giannoula2022sparsep,giannoula2022sigmetrics,denzler2021casper,boroumand2021polynesia,boroumand2021icde,singh2021fpga,singh2021accelerating,herruzo2021enabling,yavits2021giraf,asgarifafnir,boroumand2021google,amiraliphd,seshadri.bookchapter17,diab2022hicomb,fujiki2018memory,zha2020hyper,mutlu.imw13,mutlu.superfri15,ahmed2019compiler,jain2018computing,ghiasi2022genstore,deoliveira2021IEEE,cho2020mcdram,shin2018mcdram,gu2020ipim,lavenier2020,Zois2018, upmem,upmem2018, gomezluna2022ieeeaccess,  gomezluna2021cut} for various applications using accelerators or general-purpose cores. In~\cite{drumond2017mondrian}, ARM cores are used as NDP compute units to improve data analytics operators (e.g., group, join, sort). IMPICA~\cite{Hsieh2016accelerating} is an NDP pointer chasing accelerator. Tesseract~\cite{ahy+15} is a scalable NDP accelerator for parallel graph processing. TETRIS~\cite{gao2017tetris} is an NDP neural network accelerator. Lee et al.~\cite{LMd+18} propose an NDP accelerator for similarity search. GRIM-Filter~\cite{kim.bmc18} is an NDP accelerator for pre-alignment filtering~\cite{xin2012fasthash,alser2019shouji,xin2015shifted,alser2019sneakysnake,Alser2017GateKeeper} in genome analysis~\cite{alser2020accelerating}. Boroumand et al.~\cite{boroumand2018google} analyze the energy and performance impact of data movement for several widely-used Google consumer workloads, providing NDP accelerators for them. CoNDA~\cite{boroumand2019conda} provides efficient cache coherence support for NDP accelerators. 
SparseP~\cite{Giannoula2022SparsePPomacs,Giannoula2022SparsePSigmetrics} provides efficient data partitioning/maping techniques of the SpMV kernel tailored for near-bank NDP architectures.
Finally, an NDP architecture~\cite{pugsley2014ndc} has been proposed for MapReduce-style applications.. Xu et al.~\cite{XLW+18} propose a memristor-based accelerator for accelerating the sDTW kernel. Despite promising performance, they do not discuss endurance challenges associated with memristors that restrict the lifetime of the accelerator. In contrast, \myName{} considers this challenge and offers a usable lifetime of several decades.
Chen and Gu~\cite{CG21} propose an sDTW accelerator that exploits DTW pipelining using a specially designed time flip-flop. Although this work uses memristors for computation, they do not leverage PUM. The data must be moved from/to memory (i.e., memristors do not store the data). In contrast, MATSA eliminates off-chip data movement to obtain high performance and energy efficiency.
\section{Conclusions}
This paper presents \myName{}, the first MRAM-based Accelerator for Time Series Analysis. The key idea is to exploit magnetoresistive crossbars to enable energy-efficient and fast time series computation in memory. \myName{} provides the following key benefits: 1) significantly higher parallelism exploiting column-level bitwise operations, and 2) reduction in data movement overheads by leveraging PUM. \myName{} improves performance and energy consumption over CPU, GPU, FPGA, and PNM platforms.
\section*{Acknowledgements}

This work has been supported by TIN2016-80920-R and UMA18-FEDERJA-197 Spanish projects, and HiPEAC collaboration grants. 
We also acknowledge support from the SAFARI Group's industrial partners, especially ASML, Facebook, Google, Huawei, Intel, Microsoft, and VMware, as well as support from Semiconductor Research Corporation.



\balance
\bibliographystyle{unsrt}
\bibliography{mergedrefs}

\end{document}